\documentclass[12pt,preprint]{aastex}

\shorttitle{UHWFI}
\shortauthors{
Klaus W. Hodapp, 
Andreas Seifahrt, 
Gerard A. Luppino, 
Richard Wainscoat,
Ed Sousa,
Hubert Yamada, 
Alan Ryan,
Richard Shelton,
Mel Inouye,
Andrew J. Pickles,
Yanko K. Ivanov
}

\begin{document}

\title{
The University of Hawaii Wide Field Imager (UHWFI)
}

\author{
Klaus W. Hodapp\altaffilmark{1}, 
Andreas Seifahrt\altaffilmark{1},\altaffilmark{3}, 
Gerard A. Luppino\altaffilmark{2}, 
Richard Wainscoat\altaffilmark{2}, 
Ed Sousa\altaffilmark{1}, 
Hubert Yamada\altaffilmark{2}, 
Alan Ryan\altaffilmark{1}, 
Richard Shelton\altaffilmark{1},
Mel Inouye\altaffilmark{1},
Andrew J. Pickles\altaffilmark{1},\altaffilmark{4}, 
Yanko K. Ivanov\altaffilmark{1}
}

\altaffiltext{1}{
Institute for Astronomy, University of Hawaii,\\
640 N. Aohoku Place, Hilo, HI 96720, email: hodapp@ifa.hawaii.edu
}

\altaffiltext{2}{
Institute for Astronomy, University of Hawaii,\\
2680 Woodlawn Drive, Honolulu, HI 96822
}

\altaffiltext{3}{
Astrophysikalisches Institut und Universit\"{a}tssternwarte Jena,\\
Schillerg\"{a}sschen 2-3, D-07745 Jena, Germany
}

\altaffiltext{4}{Present Address:
Caltech Optical Observatories,\\
Caltech, Astronomy 105-24, Pasadena, CA 91125,\\
email: pick@caltech.edu
}

\begin{abstract} 

The University of Hawaii Wide-Field Imager (UHWFI) is a focal compressor system designed 
to project the full half-degree field of the UH~2.2~m telescope onto 
the refurbished UH~8K$\times$8K CCD camera. 
The optics use Ohara glasses and are mounted in an oil-filled cell 
to minimize light losses and ghost images from the large number of internal lens surfaces. 
The UHWFI is equipped with a six-position filter wheel and a rotating sector blade shutter, 
both driven by stepper motors. 
The instrument saw first light in 2004 in an engineering mode.
After filling the lens cell with index matching oil, integration of all software
components into the user interface, tuning of the CCD performance, and the purchase of
the final filter set, UHWFI now fully
commissioned at the UH~2.2~m telescope.

\end{abstract}

\keywords{Instrumentation: imagers -- instrumentation: detectors}

\section{Introduction}

The University of Hawaii (UH) 2.2~m telescope on Mauna Kea is a 
Ritchey-Chretien optical system originally designed in the 1960s to image large fields 
in seeing-limited quality onto photographic plates. 
While wide-field imaging is one of the main functions provided by small 
and mid-sized telescopes in today's era of large (8-m class) telescope facilities, 
the wide field capability of the UH telescope has not been 
fully utilized 
since photographic work was effectively abandoned 
some 20 years ago. 

The desire to use the UH~2.2~m telescope in a flexible schedule for a variety of search and
monitoring programs
lead to the building of the UH~8K$\times$8K~CCD camera,
which was the first large mosaic CCD camera in the world \citep{lup95}. 
This original version of the UH 8K camera provided 
an 18$\arcmin\times$18$\arcmin$ field at the f/10 Ritchey-Chretien focus of the 
UH~2.2~m telescope using a field flattener lens as the dewar window.
The UH~8K camera was also used at the CFHT~3.6m telescope prime focus but has since 
been replaced there first by the CFHT~12K camera \citep{sta00} and more
recently by Megacam \citep{bou03}. 

The UH 8K camera was recently upgraded with science-grade deep depletion CCDs.
These new, backside illuminated
and anti-reflection coated deep depletion CCID-20 devices give the UH 8K camera 
a substantially increased sensitivity per pixel, in particular in the 
far optical (0.8~-~1.1~$\mu$m) part of the spectrum.

As an additional step toward improving the capabilities of the UH~2.2~m telescope
and toward regaining the full wide field capability of this telescope
we have built a new focal compression optical system, filter wheel and shutter.
In conjunction with this new University of Hawaii Wide Field Imager (UHWFI), 
the refurbished UH~8K camera covers essentially the full field that the
UH~2.2~m telescope provides.
This paper describes the main design features of this new system.

\section{Basic Design Choices}

We initially explored various ways to increase the field coverage of the 
8K~CCD camera for search and monitoring programs. Building an entirely new prime focus system 
for the UH~2.2~m telescope was considered as a way of obtaining fields in excess of 30$\arcmin$
but was rejected since it was prohibitively expensive and would preclude the 
rapid change between instruments. A simpler and more cost effective solution 
is to build a focal compressor and field corrector that allows the full
unvignetted field of $\approx$ 30$\arcmin$ of the telescope
field to be observed at a seeing-limited pixel scale. 

The design of a wide field corrector for the UH~2.2~m telescope was constrained
by many design features of the telescope, operational considerations, and budget limitations. 
The UH~2.2~m telescope has two bent Cassegrain ports and the standard 
direct Cassegrain focus position. Operationally, using the UHWFI at one of the
bent Cassegrain foci and leaving it installed there permanently would have been
advantageous. However, the bent Cassegrain focus has a more limited 
field of view than the direct Cassegrain focus. 
The direct focus position was therefore chosen for UHWFI.

At the Cassegrain focus, the existing autoguider limits the field of view.
We decided therefore to use UHWFI without the autoguider.
For short exposure times, 
we can rely on the open-loop tracking of the telescope, which has been substantially
improved with the new control sytem \citep{lov00} of the UH~2.2~m telescope. 
We have mechanically prepared the UH~8K camera focal plane assembly for
the future installation of small guide CCD chips at
the edges of the 8K$\times$8K~CCD mosaic in a planned upgrade of this CCD camera.

\section{Optics Design}

The focal compressor optics are designed to shorten the focal length of the
UH~2.2~m telescope from the original 22.9~m to a focal length of f=13.5~m.
The focal plane scale therefore is 65.4~$\mu$m/arcsec, resulting in a pixel 
scale of 0.229 arcsec pixel$^{-1}$ for the 15~$\mu$m CCD pixels of the UH~8K, 
and approximately a $31.7\arcmin\times31.7\arcmin$ field of view, taking into account some 
dead space between the individual CCDs in the 8K$\times$8K mosaic focal plane. 
The extreme corners of the field are 
slightly ($<$10\%) vignetted by the secondary mirror, but the vignetting 
is well within what is easily correctable by proper flat-fielding of the data. 
Grid distortion by the focal compressor is a maximum of 2.8\% in 
the corners of the field and can be corrected in software, for
example using the IRAF ''geomap'' and ''geotran'' tasks.

The optical design was optimized using an as-built model of the UH 2.2 m telescope
optics. The slight spherical aberration caused by the mismatch of the primary
and secondary mirror conic constants was included in the optimization process
of the lenses and could therefore be partly corrected.

The design of the optics for UHWFI was an iterative process,
adjusting the design several times to the availability and price
of optical glasses. As a result of this process, the design represents
a relatively low-cost solution.
All designs considered in this process 
consisted of a first lens group with positive optical power, doing the actual
focal compression,
and a field-flattening lens of negative optical power close 
to the focal plane. The latter also served
as the CCD camera dewar window. The size of the first and largest lens is
limited by the diameter of the Cassegrain port in the telescope's primary
mirror cell.
For cost reasons, we were forced to abandon very good triplet designs for the first
lens group that 
used two large CaF$_2$ lenses in combination
with a single optical glass lens of negative power. With a lens diameter of 32~cm and a thickness
of 7.5~cm for the largest CaF$_2$ lens, the cost of the CaF$_2$ material was
prohibitive for our project. 

Our design now substitutes 
Ohara S-FPL51Y glass for CaF$_2$. This glass is 
the lowest dispersion glass in Ohara's product line that could be 
fabricated in the diameter required for UHWFI. 
Since S-FPL51Y cannot be fabricated in thick plates due to 
crystallization problems during the cooldown, the two thick lenses of the original 
CaF$_2$ design both are split up into two thinner ones that are within the 
fabrication limit of the raw S-FPL51Y material. 
Also, as shown in Fig. 1, we are now using two different glasses between the 
S-FPL51Y elements in the front and back of the first lens group, for
a slight gain in chromatic performance over the use of a single glass.

NOTE: Fig. 1 should be placed here.

NOTE: Table 1 should be placed here.

The first lens group thus consists of 6 lenses, four of which are Ohara glass S-FPL51Y. 
The second group is an individual lens that 
combines the functions of a field flattener and the dewar window.
A problem with this design approach is the large number of internal
surfaces (10) in the first lens group. 
Anti-reflection coatings on all air-glass interfaces could have
mitigated, but not completely eliminated the double-reflection ghost images, added
scattered light and overall loss of throughput caused by reflections on these
10 surfaces.
Instead, our design leaves the internal surfaces in the 6-lens group
uncoated and immerses these internal surfaces in a refractive-index-matching 
oil. Since the 6-lens group involves three different
optical glasses, it is impossible to choose an oil that eliminates
all internal Fresnel losses completely, but the choice of oil can be optimized
to maximize the overall throughput.

Oil spaces in
optical systems must be narrow to avoid problems with the strong
temperature dependence of the refractive index of liquids that in large
volumes can lead, effectively, to thermally induced stria. On the other
hand, the spacing between the lenses must be sufficiently large to avoid
differential thermal expansion between the different glasses 
from bringing the curved glass surfaces into direct
contact at any temperature that the system will likely encounter. In the case
of our choices of optical materials, the coefficients of thermal expansion
vary by about a factor of 2 between S-FPL51Y and the other 
glasses (S-LAL10 and S-BSM81) in the lens group.
Therefore,
our design has matching radii on adjacent surfaces, and uses 50~$\mu$m
shim stock spacers to keep the lenses at a controlled spacing large
enough to avoid them touching under any realistic temperature condition.

For the lens materials used in the main lens group, we use a refractive index
liquid from Cargille Laboratories with n$_D$~=~1.570$\pm$0.0002 at T~=~25$^{\circ}$C
that is available off-the-shelf as part of their ''A'' series of refractive
index liquids. At the average nighttime ambient temperature on Mauna Kea ($\approx$0$^\circ$C),
the refractive index changes to n$_D$~=~1.580. The liquid is transparent over the
wavelength range used for UHWFI (0.4~-~1.05~$\mu$m). At a wavelength of 0.405~$\mu$m, the transmittance
of the liquid is 97.7\% for the pathlength used in our design, and reaches 99.8\% at 0.484~$\mu$m. 
While this liquid is specified for storage and use at room
temperature, we found no problem with filling the lens cell at temperatures
near 0$^\circ$C on Mauna Kea.

The front surface of the first lens and the back surface of the last
element in the 6-lens group, as well as both surfaces of the field
flattener lens are broad-band anti-reflection coated and achieve a reflectivity
of less than 1\% from 0.44~$\mu$m to 0.82~$\mu$m.

The optical design was optimized for the 0.45~-~1.0~$\mu$m wavelength range
with particular emphasis on the longer wavelengths, to offer the best
optical performance where the new deep-depletion CCDs have their strongest competitive
advantage. 
Over this range, the chromatic aberrations of the lens system 
are not significant relative to the typical seeing so that 
very broad filters can be used for search programs. 
Fig. 2 shows a geometric spot diagram of the optics, at wavelengths from
0.4~$\mu$m to 1.0~$\mu$m. Some of the 0.4~$\mu$m spots are outside of the 3 pixel ($\approx$0.7$\arcsec$)
box, while for the other wavelengths, the spots are all within that box.
Fig. 3 shows the rms spot sizes integrated over the bandpasses of the ''grizy'' filters
listed in Table 2. The performance in the ''g'' filter is not as good as
in the other filters of our broad-band filter set, due to the poor
performance of the optics at 0.40~$\mu$m.

The optics design was optimized for 0$^{\circ}$C and 0.6~atm ambient atmospheric pressure,
the average conditions on Mauna Kea. 
The performance of the optical system remains well within 
specification over the expected temperature range on Mauna Kea ($\approx$ -5$^\circ$C to +5$^\circ$C). 

The lenses were fabricated by Janos Technology, Inc. 
The typical lens specifications were a diameter tolerance of 0.1 mm, center thickness
tolerance of 0.15 mm, edge thickness variation of 0.025 mm, and surface polish of 40/20.
Surface irregularities of two interference fringes (at 0.63~$\mu$m) over the full clear
aperture of 93\% of diameter against a testplate were acceptable. Small-scale irregularities
were specified to be
less than 1/6 fringe over any interferometer aperture of typically 80 mm diameter, the size
of interferometer available to the lens manufacturer.

NOTE: Fig. 2 should be placed here. 

NOTE: Fig. 3 should be placed here. 

\section{Mechanical Design}

The housing of the UHWFI (Fig.~4) mounts directly to the Cassegrain
rotator flange of the UH~2.2~m telescope (Fig.~5). On the top side of the housing, the lens
cell is mounted and protrudes as far into the primary mirror baffle
cone as is possible without substantial vignetting of the light path.
On its bottom side, it interfaces to the refurbished dewar of the UH~8K 
camera, whose dewar window is the last element of our optical system, the field flattener lens.
The housing contains the filter and shutter wheels and their motor and
drive gears, as well as access ports for filter changes and the installation
of the polarizer rotator HIPPO.

The UHWFI has five major mechanical components:
\begin{itemize}
 \item The focal reducer main lens group oil filled cell
 \item The filter wheel and rotary shutter assembly
 \item The interface to the 8K dewar including the field flattener lens that
serves as the dewar window
 \item A side access port for the HIPPO polarizer unit
 \item The refurbished 8K CCD camera dewar
\end{itemize}

NOTE: Fig. 4 should be placed here.

NOTE: Fig. 5 should be placed here.

While the position of the filters is not in itself very critical, the filter 
and shutter wheel housing must be very stiff to support the lens mount and the 
dewar within tight tolerances imposed by the optical power of the 
field flattener dewar window. The housing for the filter and shutter wheel
is fabricated from two thick aluminum plates, milled out to provide 
space and attachment points for the two wheel mechanisms, and also for weight
reduction. An essential component in our design is the hollow
central post in the filter and shutter wheel housing around 
which the co-axial filter and shutter wheels rotate on large diameter
Kaydon ''Really Slim'' X-contact ball bearings. 
This post is the only rigid support for the UH~8K camera on the side of the optical axis 
facing the wheels (left of the lens assembly in Fig. 4). Without this central support, 
an excessive wall thickness for the wheel housing would have been required to
meet the rigidity requirements.

\section{Detailed Design of Oil-Filled Lens}

The lens cell relies on the proper centering of the lens outer 
diameters relative to the optical surfaces without any provisions for
adjusting the centering of the lenses in the cell. 
This design approach required a 25~$\mu$m centering 
precision from the lens manufacturer, which was achieved. 
The differential shrinkage of aluminum relative to the glasses used 
between the fabrication temperature (20$^o$C) and the operating temperature (0$^o$C) 
is of the same order as the alignment tolerances and was accounted for in
the design of the lens cell. 
We are using a design similar to that used for cryogenic lenses 
\citep{hod03}
and support the lenses radially against two hard reference surfaces 
using an opposing spring loaded plunger
to maintain contact with the reference surfaces
in any orientation of the instrument. Prior to applying the spring preload, 
the lenses are not tightly constrained radially, 
so that they can be easily installed. 

Axially, the lenses are spaced by thin (50~$\mu$m) spacers, as explained
above and as seen in Fig.~5.  
Both sides 
of the lens stack are axially loaded by compressed flexible Viton O-rings that 
also serve to seal the immersion oil cell.
This arrangement avoids tight fits during assembly of the lenses 
and minimizes the potential for damaging the lenses, in particular the
very fragile S-FPL51Y lenses, during installation.

The focal compressor main lens group is assembled by first 
stacking the lenses upside down on the front
ring of the lens assembly (Fig.~6). 
During the initial stack-up, the lenses are only 
roughly centered. After stacking, the body of the lens cell is lowered over
the stack and closed, but not yet axially compressed. The lens cell is then
put on its side with the two radial hard points on the bottom, so that the individual
lenses will slide by $\approx$ 1 mm into contact with the radial support
points. Next, the preload springs are installed
and preloaded to keep the lenses positioned radially. 
Finally, the axial lens cell screws
are tightened to compress the O-ring, which then define the lens positions axially and
provide the main oil seal. Finally, the radial
preload screw ports are sealed. 
The compression of the O-rings was measured and the spacing between the 6-lens group
and the field flattener lens was adjusted in a final machining step 
to give the proper spacing of these two
components.

NOTE: Fig. 6 should be placed here.

The lens system can be used in this dry condition, even though without the index
matching oil, ghost imaging is more severe and the throughput is lower.
Only after a successful test run at the telescope and full verification of
the optical performance did we flood the cell with the index-matching oil.

NOTE: Fig. 7 should be placed here.

Fig.~7 shows the lens cell during the oil filling procedure. 
The optical axis is pointed horizontally.
The index matching oil is being supplied from the lowest side of the cell and
is drawn up into the lens gaps by capillary force. The supply of
oil is adjusted for a very slow fill, taking about 20 hours to complete,
so that the accidental inclusion of bubbles can be avoided.
The image shows the lens about half filled. The reduction in the number
and intensity of reflections and the substantial increase in overall
throughput is clearly visible.

When the lens is filled with enough oil to fully penetrate the 
inter lens spaces, a bubble of air is left in the remaining volume
of the lens cell. This compressible air volume is important to
accommodate changes in the oil volume due to temperature changes.
While this air bubble is free to move within the lens cell in
response to changing orientation of the instrument, the surface tension
at the oil-air interface prevents this large bubble from entering the narrow
(50~$\mu$m) spaces between the lenses.

\section{Filter and Shutter Wheel Design}

The filter wheel in UHWFI carries six filters of dimension 165$\times$165~mm and up to 15~mm thickness;
the nominal filter thickness in UHWFI is 10 mm. 
This large wheel is driven by a Geneva drive conceptually similar 
to those developed by \citet{bel98} for some of the cryogenic instruments built 
at the IfA. 

The advantages of this design are that it can be fully 
fabricated in-house, that it allows loose tolerances between 
the wheel and the drive, and does not pose difficult motor control requirements. 
Used in conjunction with a spring-loaded detent, it nevertheless achieves very 
high positioning precision. Similar again to the cryogenic 
mechanisms designed for other instruments, the wheel position 
is encoded by small magnets inserted into the wheel, 
and their magnetic field sensed by Hall effect sensors 
on the wheel housing.  To facilitate the exchange of filters, 
the filters are individually mounted in cassettes 
that allow their handling without contact to the optical surfaces. 
The filter wheel housing has an access port (Fig.~4 and Fig.~5) so that 
filters can be exchanged easily while the instrument 
is mounted at the telescope.

The UHWFI is equipped with a newly designed 
large (diameter 71 cm) rotary sector blade shutter.
The shutter blade is made of a lightweight
aluminum honeycomb disk with outer diameter of 71 cm 
and inner diameter of 24 cm with a single machined 120$^\circ$ sector cutout.
This design results in a low inertia, but still self-supporting blade.  
The shutter sector wheel is carefully balanced with
a counterweight.

The shutter wheel  
is co-axial with the filter wheel. 
Stacking these two large wheels on top of each other 
offers significant design advantages over the combination 
of a filter wheel and a conventional linear shutter. 
The shutter is driven by a stepper motor (Phytron ZSH57) via a
5~mm pitch, 15~mm wide timing belt with a gear reduction of 11:1 under the
control of a Galil motor controller. Position feedback is via
embedded magnets in the shutter blade sensed by
Hall effect sensors (from F. W. Bell). 

The shutter is triggered by the rising edge of a TTL signal generated by the
CCD readout Leach controller.
This design was chosen for software compatibility with the existing
UH CCD cameras.
The shutter opens with a rapid acceleration phase
of the shutter blade, essentially the maximum torque acceleration
profile that
the motor can supply. The acceleration phase is completed before
any of the shutter blade edges intersect the optical path. 
While crossing the optical path, both in opening and closing
direction, the shutter blade rotates at constant angular velocity.
The shutter blade then decelerates again and comes to a standstill
in the open position. The closing of the shutter is triggered by
the trailing edge of the TTL signal from the CCD controller and
proceeds similar to the opening phase.

Due to its large size and inertia, the shutter is not designed for short
exposures. The shortest possible exposure time, limited by the
mechanical properties of the shutter
is about 1~s. The exposure times were calibrated using observations
of bright stars. Above nominal exposure times of 2~s, the effective
shutter open time is 0.25~s shorter than the nominal exposure time set
by the duration of the CCD controller TTL signal.

\section{Dewar Window}

The last lens element is a field flattener lens that is required to
compensate for the substantial field curvature of the original
Ritchey-Chretien telescope optics and of the focal compressor lens
group. To minimize the number of optical surfaces, this last lens also doubles
as the dewar window. Since is is subject to high atmospheric pressure forces, we only
support this lens axially on a (soft) O-ring that gets compressed
by several millimeters when the dewar is evacuated. 
The compression of the O-ring under atmospheric pressure of 0.6 atm
was measured and the design spacing
adjusted to give the proper spacing of the field flattener lens
relative to the focal compressor 6-lens group and the detector.

The large dewar window lens cools radiatively into the dewar when it is
cold. To avoid condensation problems on its front surface, the inner volume of 
the UHWFI between the compressor lens group and the dewar window
lens, including the filter wheel and shutter assemblies, is constantly being flushed with
dry nitrogen from the dewar boil-off.

\section{Control Electronics}

The UHWFI control electronics operates the stepper motors for the filter and
shutter wheels, and reads back home position information from the Hall effect
sensors.
It consists of a Galil Motion Control Inc. DMC-2120 2-axis Stand-Alone Motion Controller
connected through a Galil ICM-2900 Interface Module to two Phytron Inc. ZSO MINI
Bipolar Stepper Motor Driver Modules, stepper motor incremental encoders and an
IfA built multi-channel Hall effect sensor support board. 
The host computer communicates with the motion controller via ethernet to initiate shutter or filter
wheel actions or to obtain position information.  

The Phytron drive modules generate the phased pulses required to move the Phytron
Inc. ZSH 57 stepper motors from a dual output +62 V DC power supply.  The signals
from the two ZHS 57 stepper motors' integrated incremental encoders are used to
complete the DMC-2120 motion controller's feedback control loop.  The motion
controller also sets the acceleration, deceleration, and maximum speed of the
motors.

Hall effect sensors are used to determine positions by detecting small rare-earth
magnets embedded in the shutter and filter wheels.  One sensor is used to locate
the ''home'' or closed position of the shutter.  Three sensors and associated
magnets are used to uniquely identify each of the eight filter wheel positions. 
The IfA built multi-channel Hall effect sensor support board (same design as
for the NIRI instrument \citep{hod03}) provides precision
stable voltage to the sensors and to on board voltage comparators as well as gain,
filtering and conversion of the sensor signals.  A Burr-Brown INA141 instrumentation
amplifier provides gain and converts the differential sensor signals to single
ended signals.  The buffered signals are then sent to the external A/D converter
of the Galil controller.
The Hall effect sensors are used to initially find the wheel ''home'' positions.
Normal
movements of the shutter or filter wheel are done by driving the motors with
a predetermined number of steps and 
the Hall effect sensors are used only to verify the end positions.

\section{The UH~8K CCD Camera}

The UH~8K camera was first commissioned in the Spring of 1995 at the
prime focus of the CFHT. The camera was initially equipped with 8
Loral/Fairchild 2K$\times$4K 15~$\mu$m pixel CCDs and offered an
image scale of 0.21\arcsec pixel$^{-1}$ and a field
of view of 0.47$^\circ$$\times$0.47$^\circ$ on this telescope \citep{lup95}. In addition,
with the use of a field flattener as the camera entrance
window, the camera has been used at the f/10 focus of the UH~2.2~m 
telescope, where the image scale is 0.14\arcsec pixel$^{-1}$ and the field of 
view is 0.31$^\circ$$\times$0.31$^\circ$.

Because of budget constraints as well as our desire to deploy
this camera as early as possible in order to take advantage of some
scientific opportunities, we had elected at the time to use detectors that
would not normally be considered "science-grade."
These early detectors were operated in "front-illuminated" mode resulting
in a peak quantum efficiency of only 40\% and with
little or no response below 0.4~$\mu$m. Furthermore,
the CCDs had fairly high readnoise ($>$10~e$^-$), 
along with some serious charge transfer issues that
required that we operate at a higher-than-usual temperature
(-70$^o$C). This then caused problems with spatially varying dark current.
Finally, the readout time of the camera exceeded 7 minutes leading
to a low duty cycle operation in practice.

We began a program
to replace the UH8K camera CCDs with superb, science-grade detectors
obtained as part of the UH-led MIT Lincoln Lab Consortium. The
UH/MITLL project was initiated to design and build 2K$\times$4K 15~$\mu$m pixel
CCDs on high resistivity silicon and thin and back illuminate them
to achieve the highest possible quantum efficiency over the entire optical region, but
with particular emphasis on the red end (0.7~-~1.1~$\mu$m) where the thicker
detectors offered a real advantage over their typical thin CCD counterparts.
The CCD that resulted from this effort was called the CCID-20 and
these detectors have been incorporated into various large
instruments including Subaru SUPRIME, Keck ESI, CFHT12K,
Keck DEIMOS, ESO VLT, AAO/MSSSO WFI and UH AEOS spectrograph. They
have achieved $<$~2~e$^-$ readnoise at 100kpix~s$^{-1}$ readout speed, very good
charge transfer efficiency, and a peak quantum efficiency 
at 0.6~$\mu$m exceeding 90\% with a QE at 1~$\mu$m of 20\%.

NOTE: Fig. 8 should be placed here.

We elected to upgrade the UH~8K camera following our experience building the
nearly identical CFH12K \citep{sta00}. We rebuilt the cryostat and the focal plane (Fig.~8),
and acquired new faster San Diego State University (SDSU2) controllers, identical
to those used in CFH12K. With these improvements,
we expected to see a substantial gain in overall performance:
7x reduction in readout time, a reduction by a factor of 3-5 in readout noise,
and, on average, a factor of 2 improvement in quantum efficiency, with a
larger improvement at the long wavelength end. 
Also, the operating temperature of the new CCDs is lower and therefore,
dark currents should no longer be an issue.

NOTE: Table 2 should be placed here

The UHWFI is now equipped with a set of broad-band g, r, i, z, and y filters
closely matching those planned for the Pan-STARRS project, as well as narrow-band
H$\alpha$ and [SII] filters. In addition, a wide V+R filter is being used for
asteroid searches.

After the refurbishment of the UH~8K camera with new CCDs and a new
set of readout electronics, the
readout time is now 60 s and the total overhead over the shutter open
time is 63 s per exposure. Quantum efficiency is greatly improved,
as are the cosmetics of the devices and the dark current.
The UH~8K camera is prepared for the installation of additional
guide CCDs, if the need for such guiding becomes important, at this time,
however, we are relying of open-loop tracking of the telescope.

The UH~8K user interface is a version of the ''detcom'' detector controller
software and its ''director'' command line user interface originally developed
for the CFH12K camera \citep{sta00}. The control of the filter wheel and shutter
wheel was integrated into the ''director'' user interface.

At this time, we don't have any guiding of the telescope
during the exposure. However, with the new telescope control system,
the UH 2.2~m telescope tracks well enough for exposures of up to about 5 minutes.
The image quality achieved by UHWFI is limited by the performance of the telescope
optics. These suffer from complex aberrations that have a strong component of astigmatism.
In practice, most observers focus the telescope to the best compromise focus where the
images appear nearly round, but are slightly blurred beyond what seeing would produce.
The telescope focus depends on temperature and telescope position, due to imperfections
in the mirror support. In practice, focus adjustments are required about every 15 minutes.
Also, astigmatism produced by focus errors is often hard to distinguish from telescope tracking imperfections.
In combination, these characteristics of the UH 2.2~m telescope force observers to 
choose relatively short integration times.
In addition, the thick deep depletion devices now used in UHWFI have a much higher cross section for
cosmic rays, and therefore cosmic ray hits begin to limit the integration
times.  In practice, most observers use integration times of 3 to 5 minutes.
This is enough to be sky background limited in all broad filters 
and to be marginally sky background limited in the H$\alpha$ and [SII]
filters. 
In such short exposures, and with proper attention paid to focus control, we are
typically achieving image quality of $\approx$0.75$\arcsec$ FWHM.

\acknowledgements 

This project was supported by NSF grant AST00-96833.

\clearpage
\begin{deluxetable}{cccccccc}
\tabletypesize{\scriptsize}
\tablecaption{UHWFI Optical As-Built Data at 20$^\circ$C}
\tablewidth{0pt}
\tablehead{
\colhead{Optical Element} & \colhead{Material} & \colhead{Thickness} 
& \colhead{Radius-1} & \colhead{Conic C.} & \colhead{Radius-2} 
& \colhead{Diameter} & \colhead{Distance to next}
\\
\colhead{Description} & \colhead{} & \colhead{of Element} & \colhead{of Curvature}
& \colhead{Constant} & \colhead{of Curvature} & \colhead{Element}
}
\startdata
Primary Mirror & Mirror & 0 & -12345 mm & -1.0535 & & 2262 mm & 4640 mm \\
Secondary Mirror & Mirror & 0 & -4198 & -3.604 & & 556 mm & 5012 mm \\
Lens-1 & S-FPL51 Y & 39.0 & 313.7 & 0 & 840.5 & 320 mm & 0.05 mm\\
Lens-2 & S-FPL51 Y & 39.0 & 840.5 & 0 & -1200 & 318 mm & 0.05 mm\\
Lens-3 & S-LAL10 & 15.0 & -1195 & 0 & plano & 316 mm & 0.05 mm\\
Lens-4 & S-BSM81 & 15.0 & plano & 0 & 325.0 & 314 mm & 0.05 mm\\
Lens-5 & S-FPL51 Y & 30.0 & 325.0 & 0 & 840.5 & 296 mm & 0.05 mm\\
Lens-6 & S-FPL51 Y & 30.0 & 840.5 & 0 & -2195 mm & 176.1 mm\\
Filter & BK7 (nominal) & 10.0 & plano & 0 & plano & 26.2 mm\\
Lens-7 & S-TIM2 & 15.00 & -3000 & 0 & 325.0 mm & 32.8 mm\\

\enddata
\end{deluxetable}

\clearpage
\begin{deluxetable}{ccc}
\tabletypesize{\scriptsize}
\tablecaption{UHWFI Filter Bandpasses}
\tablewidth{0pt}
\tablehead{
\colhead{Bandpass} & \colhead{$\lambda_1$} & \colhead{$\lambda_2$}
}
\startdata
g & 402 & 552 \\
r & 552 & 691 \\
i & 691 & 818 \\
z & 818 & 922 \\
y & 948 & 1060 \\
H$\alpha$ & 652 & 661 \\
$[SII]$ & 667 & 677 \\

\enddata
\end{deluxetable}

\clearpage
\begin{figure}
\plotone{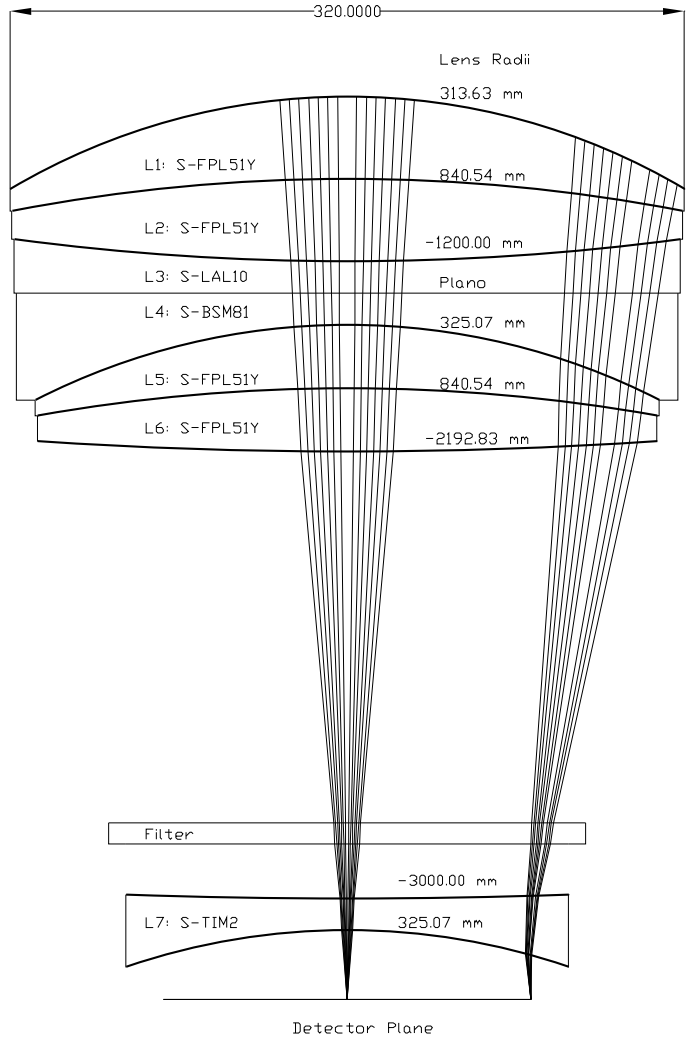}
\caption{
Cross section layout and raytrace of the UH Wide Field Imager focal compressor optics. 
The lens materials and the design radii of the lenses at the fabrication temperature 
of 20$^\circ$C are shown in the figure.
}
\end{figure}

\clearpage
\begin{figure}
\plotone{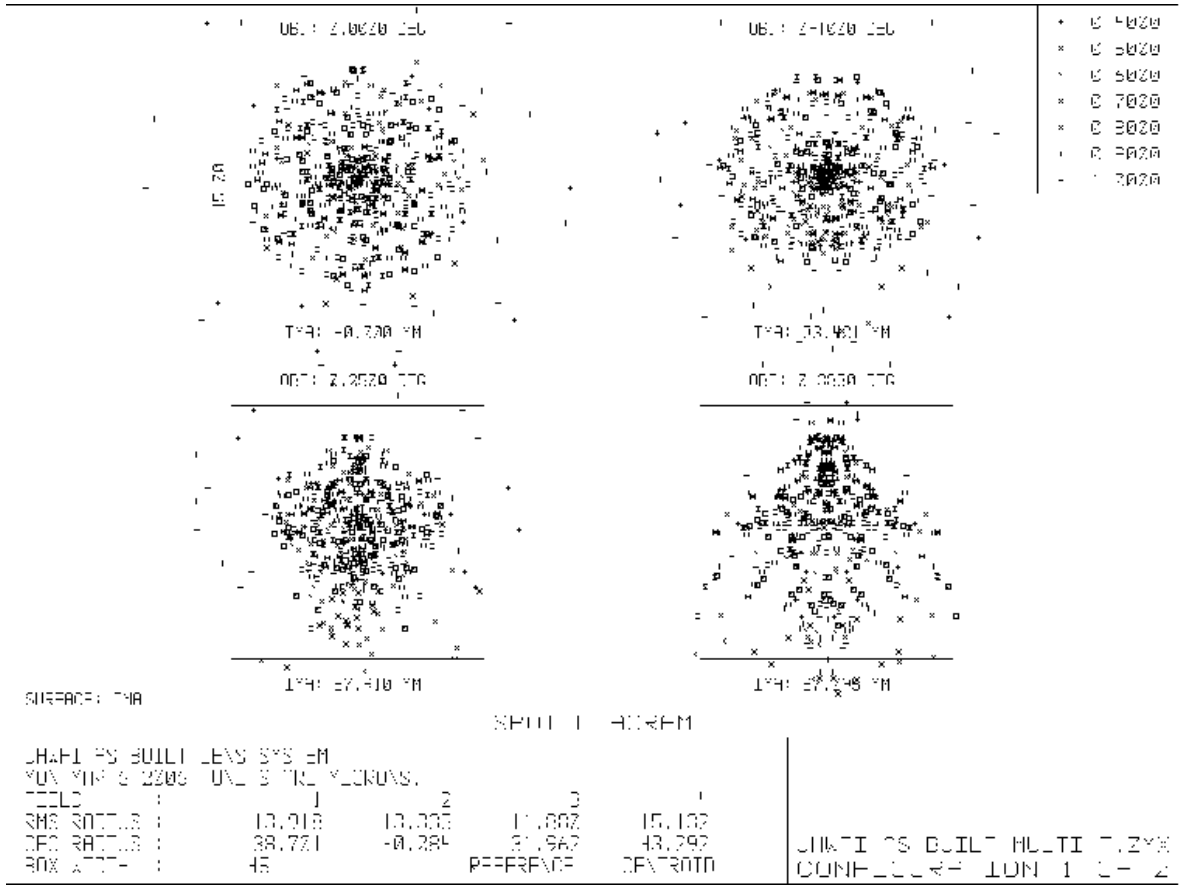}
\caption{Geometric
spot diagram of the polychromatic image quality of the 
UHWFI between 0.4~$\mu$m and 1.0~$\mu$m. 
The box represents 3 pixels (15~$\mu$m each), or $\approx$0.7$\arcsec$ on the sky,
roughly the median seeing FWHM achieved at the UH 2.2 m telescope. The four
boxes are at radial distances 
of 0.0$\arcmin$, 6.0$\arcmin$, 15.0$\arcmin$, and 23.0$\arcmin$ (field corner)  from the
optical axis.}
\end{figure}

\clearpage
\begin{figure}
\plotone{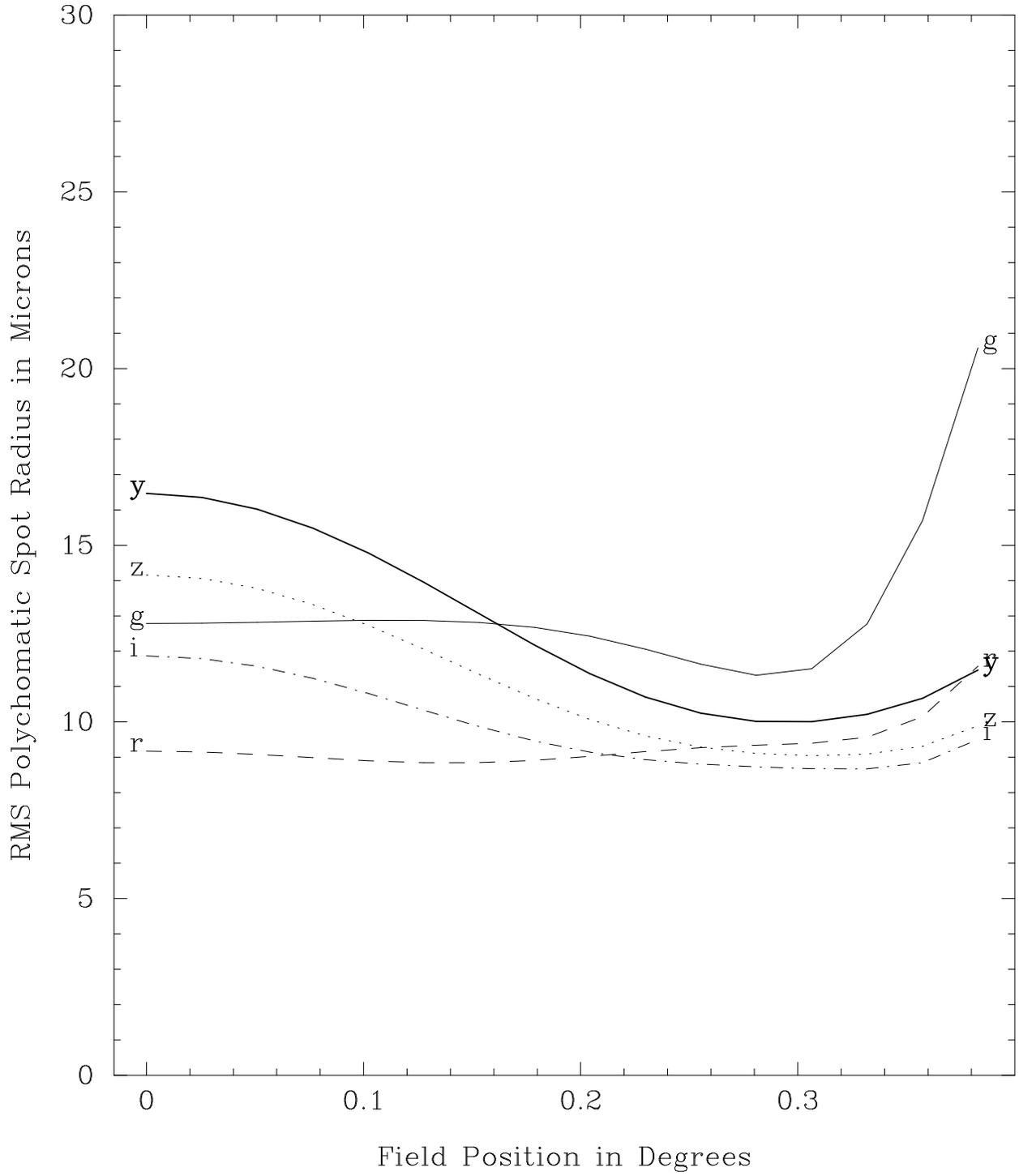}
\caption{
Geometric RMS spot size integrated over the bandpass of each filter, plotted against
the field angle.
}
\end{figure}

\clearpage
\begin{figure}
\plotone{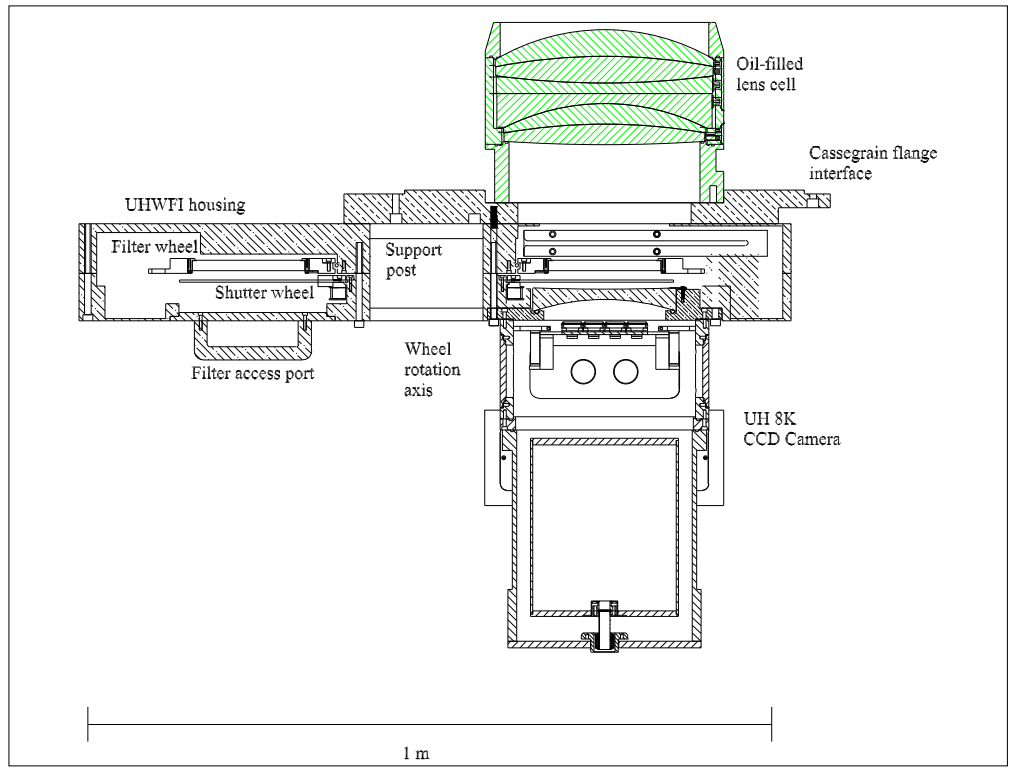}
\caption{
Mechanical Layout of the UH Wide Field Imager, 
including the lens mount, shutter mechanism, filter wheel mechanism, 
and the upgraded UH 8K dewar. 
}
\end{figure}

\clearpage
\begin{figure}
\plotone{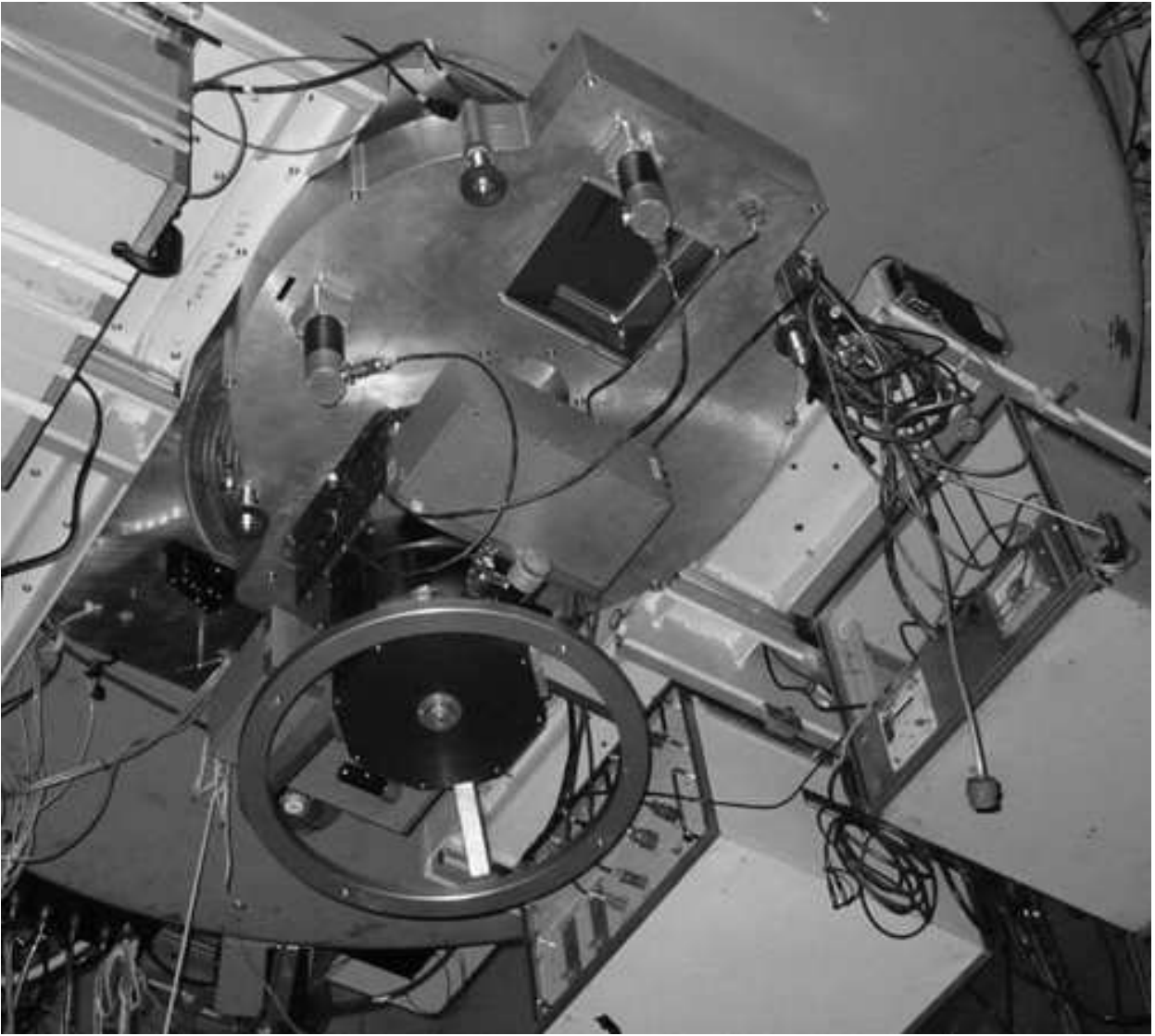}
\caption{UHWFI mounted at the UH~2.2~m telescope direct Cassegain focus.
The filter access port is open, showing an empty position of the filter wheel
and the shutter sector blade.}
\end{figure}

\clearpage
\begin{figure}
\plotone{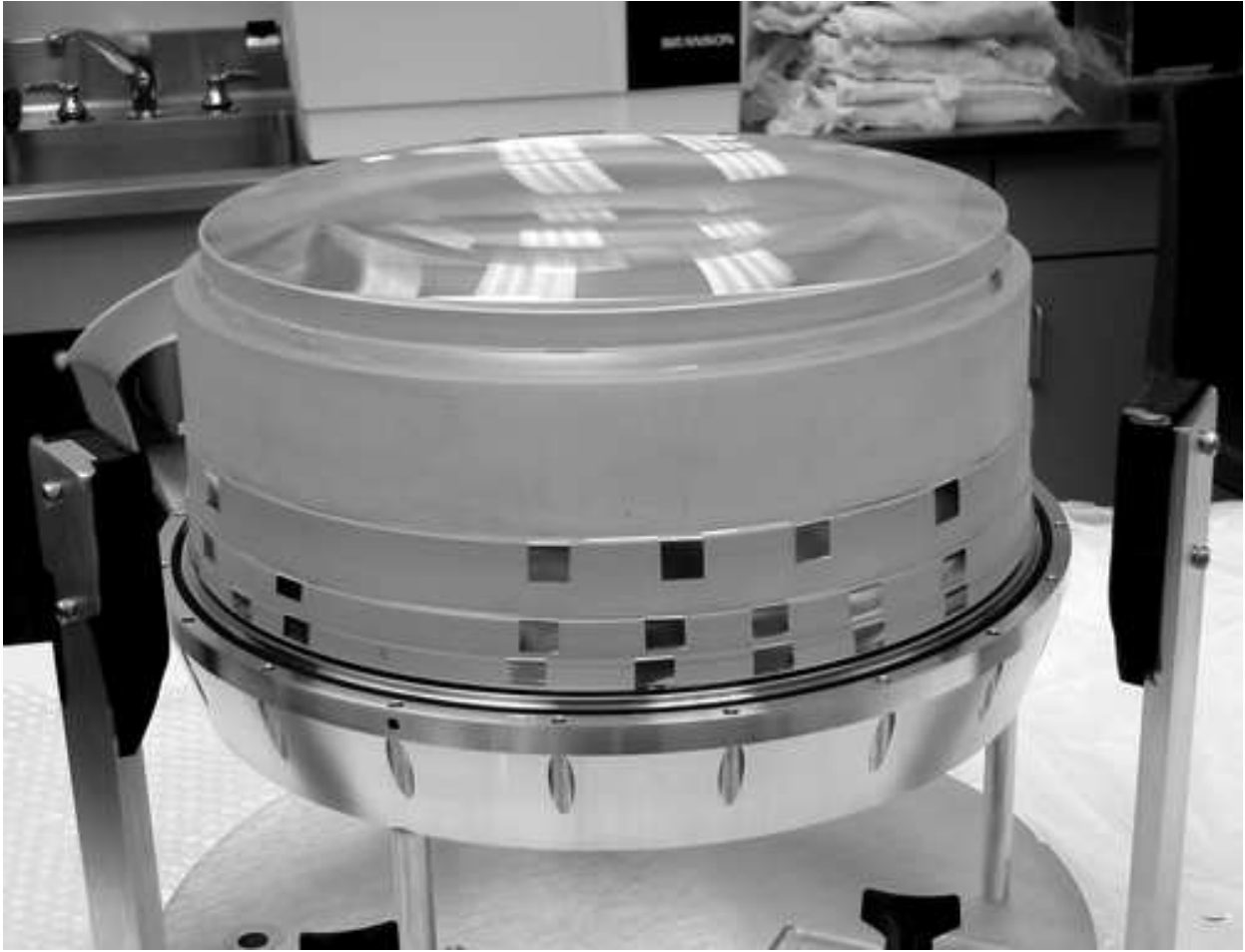}
\caption{The lenses of the 6-lens group are stacked with the front lens facing down on
the lens cell front ring.}
\end{figure}

\clearpage
\begin{figure}
\plotone{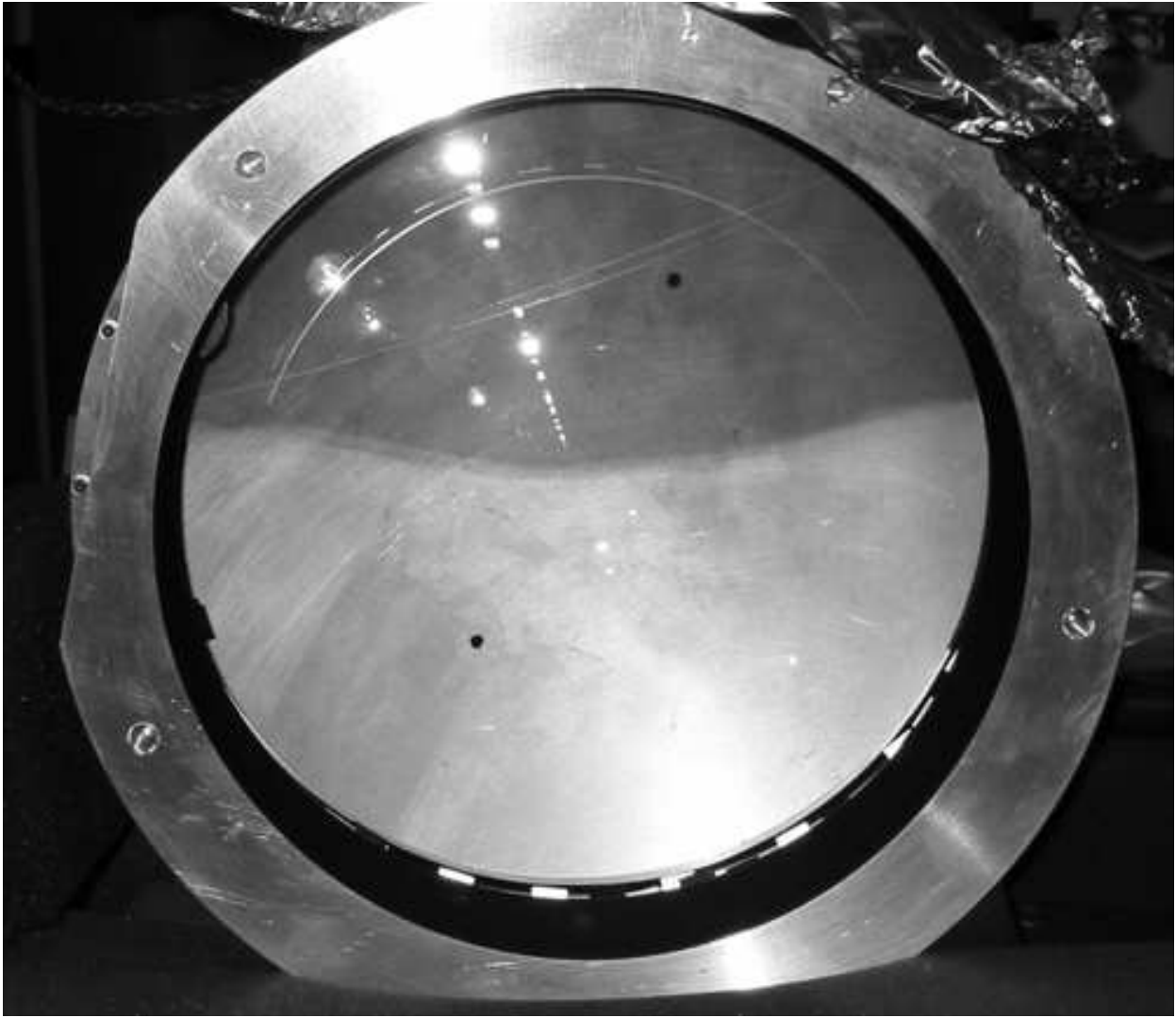}
\caption{Index matching oil is being filled in from the bottom of the side-facing
lens cell. At the time of taking this picture, the lens cell was a little over half filled.
The throughput in the oil-filled lower half is higher, and therefore, more ambient light is illuminating
an aluminum cover plate on the back of the lens cell.}
\end{figure}

\clearpage
\begin{figure}
\plotone{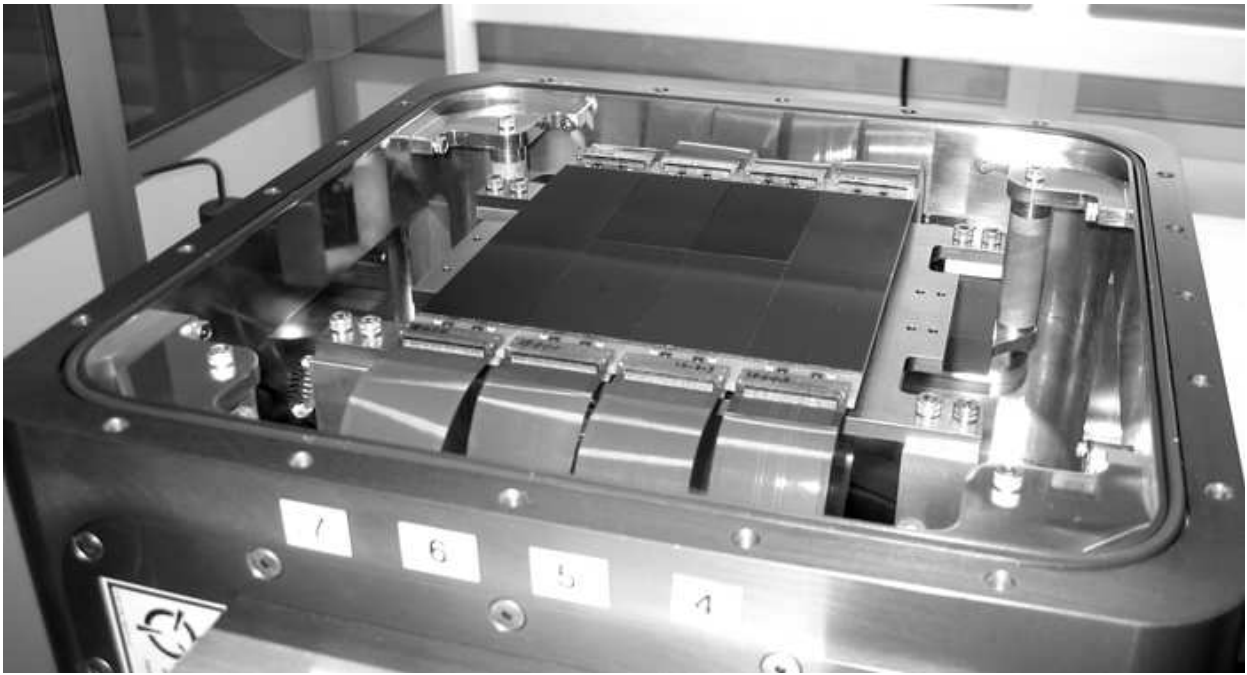}
\caption{UHWFI~8K~CCD focal plane populated with CCID-20 deep-depletion CCDs}
\end{figure}

\end{document}